\documentstyle[12pt,a4,epsf]{article}
\begin{document}
\title{Metallic Xenon.\\
Conductivity or Superconductivity?}
\author{V.N.Bogomolov\\
\normalsize \rm Russian Academy of Sciences\\
\normalsize \rm A.F.Ioffe Physical-Technical Institute\\
\normalsize \rm St.-Petersbourg, Russia}
\date{}
\maketitle

\section*{}
The metallization of gases is a problem with a long history. Suspicion that
the hydrogen under pressure to be metallic arose as far back as 19th
century [1].
In the first half of the 20th century the metallic atomic hydrogen attract
an interest as a simplest and superlight alkali metal analogy. In 1927 a
simple criterion of metallization of dielectric condensates was found [2].
In 1935 the density of atomic hydrogen was calculated and the critical
pressure (about 100 GPa) was evaluated. In 70th after the BCS  theory was
created, some ideas of non-phonon superconductivity mechanisms were
advanced. It stimulated a new splash of an interest on the hydrogen metall
ization, especially on molecular hydrogen with two electrons with the
opposite directed spins. Xenon is the only gas transformed into metal at
the present time.

Thorough investigations of the xenon metallization were carried out in
[4,5]. Some parameters of optical properties of xenon under pressure were
obtained. At  pressure larger than (107 - 130) GPa (molar volume $V_{\rm
m1}\sim$ (11.6 - 10.7) cm$^3$ /mol) appears an absorption of photons with
energy below 2.7 eV.
In Fig.1 (curve 1) the absorption coefficient of this type of absorption at
130 GPa is shown [4]. It is interpreted as an indirect interband absorption
edge. At pressure larger than 138 GPa
($V_{m2} = {\rm 10.56 cm}^3$/{mol}) a peak of absorption at 2 eV was
observed. It remains at the larger pressure, as well as the first type
absorption (Fig.1, curve 2) [4]. There is no unambiguous interpretation of
this absorption [4, 5].
A sharp rise of the absorption below about 1.5 eV (Fig. 1, curve 2 ) at the
pressure larger than 142 GPa
($V_{m3} ={\rm 10.46 cm}^3 $/{mol}) is regarded as an absorption of
classical free-electron metal near the plasma frequency $W_p$. Existence of
free electrons is a result of diminution of the energy gap under pressure
[4]. According to [4]
$$
                        W_p^{4/3} \sim (V_{m3} - V).  \eqno(1)
$$
 A summary absorption  $b$  in the peak at 2 eV may be described by the
band model relation [4]
$$
                        b^{2/3} \sim (V_{m2} - V).\eqno(2)
$$
The main purpose of these works [4,5] was to confirm the fact of
metallization but not to give its correct interpretation. Nevertheless the
high accuracy of the experimental data allows to find not only some
contradictions in the optical data interpretation, but also to consider
some alternative mechanism of molecular bonding and metallization.
In Fig 2a relations (1) and (2) are presented in a logarithmic scale.
Indexes near the straight lines are the powers corresponding to these
dependencies. It is clear that the more correct powers are 2 and 1. In Fig
2b relations (1) and (2) with correct powers are shown.
High accuracy of the experimental data and absence of their suitable
description is the cause to try to find another model of the phenomenon.
The experimental relation between the plasma absorption energy and the
molar volume difference from the volume at metallization is
$$
                  0.5W_p = 3.2 B ( 1 - V/V_{m3} ) 1/2,  \eqno ( 3 )
$$
where $B = 0.48$ eV  ( Fig 3 ).
Experimental data corresponds well to Eq. (3). This relation may describe
the energy gap near the phase transition.
Let us suppose the metallic xenon to be a superconductor. For this case
$T_c\sim 0.48\,{\rm eV} \sim 5000$\, K. The gap energy near transition is
about 1 eV. Far of the transition gap energy is about $3.5\cdot0.48
=1.68$\, eV. The temperature $T$ is replaced by the volume $V$. The
parameters $T$ and $V$ are equivalent in the Gibbs function. For the
metallic xenon $T = \rm const$, while for
ordinary superconductors $V =\rm const$.

Optical absorption by superconducting films has been carefully investigated
[6]. A theoretical frequency dependence of a transmission curve [6] may be
transformed into a frequency dependence of an absorption coefficient. In
Fig.1 the dependence of the theoretical value of the absorption coefficient
on the reduced photon energy $W/(3.5kT_c)$ is shown ( curve 3 ).

One may conclude from comparision of the curves 2 and 3 that the gap energy
for the experimental curve 2 is 1.4 eV and $T_c\sim~ 0.4$\,eV. It is close
to the energy gap evaluated from the plasma frequency. Increasing of the
absorption in a high energy region is due to the absorption by normal
electrons, as for ordinary superconductor films. For the last the hump
corresponding to 2 eV peak were observed sometimes too.

The $P\! -\! V$ dependence is plotted in Fig.4 (experimental points, curve
1) [4]. The pressure dependence of the compressibility is shown by the
curve 2 in arbitrary units. An approximation by two straight lines of two
parts of this curve is shown as well. The possible jump in compressibility
at the intersection of lines
($V_{m3}=10.46$\,cm$^3$/mol ) is neglected.

   The $P-V$ dependence with a break is given also in [5] with out comment.
Softening of the system is a result of the metallization. The touch line 3
is an extrapolation of the dielectric compressibility into the
metallization region. One may suppose the metallization not to be happened
and $P-V$ curve runs along the line 4 up to 200 GPa. At this pressure we
shall "swich" a metallization transition and find ourselves at the line 1.
Volume of the system decreases by 0.12 cm$^3$/mol and its energy changes
by 0.24 eV. It means that the number of pairs is about 0.2 of the number of
xenon atoms, because the energy gap is about ( 1--1.4) eV. System is indeed
near the phase transition.
   One of alternative description of the IG atom condensates could be based
on an average interatomic interaction, for example, the Lennard-Jones
potential ({\it6-12}). It could lead to loss of some condensate properties.
Because of the weekness of interatomic interactions electronic exitations
must be regarded as localized but not Bloch waves. It makes applicability
of band theory description of electronic properties of molecular
condensates doubtful.

The similar situation is well known in the case of small radius polarons.
In this case localizes an electron at a site much faster than spreads over
the lattice. And the electron transport is a diffusion process.  It was the
main difficulty in a description of the electronic properties of TiO$_2$ ,
treated as a narrow band material [7]. Most of the properties of TiO$_2$
was adequate described when the small radius polaron approach was utilized
[8]. It is possible the same situation take place for condensa
tes of IG atoms.
   The most typical feature of IG condensates is the smallness of atom size
$2r_1$ relatively to equilibrium interatomic distances $2r_2$ in
condensates. For xenon $2r_1 = 1.2$ \AA, while
$2r_2$ = 4.4 \AA. The bond energy of condensate is about 0.13 eV. It is
negligible in comparision to the first excitation energy of atoms
(10 eV).

  The second very important circumstance is the coincidence of interatomic
distances $2r_2$ with diameters of atoms in an excited state. Such atoms
are excimer analogies of the corresponding alcaly metals. For this case an
alternative approach to a description of properties of molecular
condensates is possible. These substances have a bond of a metallic type
via excited state orbitals but without a metallic conductivity because a
mean number of electrons at excited state orbitals $X < 1$. This situation i
s shown chematically in Fig. 3
(insertion). For xenon $X = 0.038$. Most of the physical properties
(condensation and adsorption energy, compressibility, metallization under
pressure) may be described by expressions of the theory of simple metals
with electron charge $e$ replaced by $eX$ [9]. Nevertheless such averaging
in some cases is unfit also.

 It is the fact that condensed xenon at normal pressure has
$X = 0.038$ and the electron concentration  $n = 10^{21}$
cm$^{-3}$ what corresponds to 1/25 of the alcaly metal electron
concentration. Absence of the conductivity and the low bond energy at high
electron concentration is a result of averaging at $X < 1$. Actual meaning
of $X$ is the probability for an electron to appear on an excited state
orbital. It is a mean number of virtual excimer metallic atoms among the
whole atoms of the condensate. The conductivity is absent if  $X < X_p =
0.12$ --- the percolation threshold [9].

In Fig. 3a in the lattice time scale such situation is shown. And in the
electronic time scale an instantaneous electron distribution of the virtual
molecular excitation gas is shown in Fig. 3b. It is obvious that electronic
conductivity is possible if $X > X_p$ and the band theory description of
the system is valid at $X$ much larger.

After $X$ becomes larger $X_p$ a cobweb of conducting chains or clusters of
atoms in virtually excited state penetrates a whole dielectric condensate.
Such conducting chains are surrounded by dielectric media in a pre-excimer
state. Configuration of the conducting chains depends on the atomic wave
function type. It changes permanently and statistically remaining the
percolation conductivity up to appearance of the influence of the
regularity of the lattice sites. At $X = 1$ condensate becomes excimer alca
ly metal.

An intermediate situation when the nano-dispersed metal-dielectric system
is realized is the most interesting one. It may be concerned with HTSC
problems and various constructions discussed actively in 70th.

Properties of the condensate are determined by the excited state radius
$r_2$ of atoms. It may be expressed through energy by the hydrogen-like
formulae $2r_2 =e^2 /( E_0 - E_l)$. Here $E_0$ - the ionization potential,
$E_l = e^2 / 2r_l$ --- the transition energy between the ground and excited
state of atom.

For the system in which the excited state radius of atoms is fixed the
atomic wave functions may be expressed by the energy. Probability $X$ for
an electron to appear at excited state orbital with radius $r_2$ may be
obtained from the atomic wave functions of excited state
$$
X(r_2)=X_1\exp(-r_2/r_1)=X(E)=X_2\exp(-E_1/w)=X_2\exp(-1/g).
\eqno(4)
$$
The pre-exponential coefficient is a relatively weak function of $r$.
An average perturbation energy $w=e^2/2(r_2-r_1)$ is the
interaction energy between electrons of the ground state orbitals of the
neighbouring atoms at distance $2r_2$ in condensate. For IG relation $g =
w/E_1 \sim ( 0.30 - 0.55)$. Metallization of xenon occurs at $g = 0.75$.
The main properties of condensates may be expressed through atomic
spectroscopic parameters of atoms.

As the interatomic interaction is a pair one it leads to the possibility of
the H$_2$-like molecule Xe$_2$ creation. Such virtual excimer molecules
Xe$_2$ has a bond energy of about 1 eV. It is about  $2.2/0.529$  less the
bond energy of H$_2$ molecule (4.37 eV).
Lowering of the excitation process energy by 1 eV may facilitates
generation of virtual molecular electron pairs and their participation in
both the bonding of condensate and the conductivity at the metallization.
Energies of the same order have appeared in a discussion of the metallic
xenon properties.

The conclusion is that the energy band theory either is not applicable for
a description of the metallic xenon properties or it must be more carefully
worked up.

  At the above consideration the condensation of xenon atoms is a result of
atoms interaction with participation of virtual excitations with
concentration $X$.. The conductivity arises at $V_m$ corresponding to
transition over the percolation threshold
$X_p(W) = 0.12$, which might depend on the light frequency. Fig. 5 on the
base of data [4] shows dependencies of the absorption coefficient ( the
electron concentracion ) at frequences 1 ,  1.7 ,  2 , 2.7 eV on the
difference $V_m - V$. It is clear, that the lower pressure (more $V_m$),
the more light frequence. The straight line  a  is the dependence of  $1 -
V/V_m\sim (X_p - X )$ on $W^2$. The experimental data $V_m$ correspond well
to this dependence. To a zero frequency corresponds $V_{m0} = 10.28$ cm
$^3$/mol , which is nearer to 10.2 cm$^3$/mol [2],
than $V_m = 10.7$ cm$^3$/mol [4] obtained at nonzero frequency. Up to
$V_{m3} = 10.46$ cm$^3$/mol light is absorbed by normal electrons of
metallic chains ($W_p\sim 10$ eV). But the mean concentration  $n \sim (X -
X_p)\sim(1 - V/V_m)$  (4). A peak at 2 eV might be a "precursor" of the
superconductivity, which arises at the pressure higher than 142 GPa . The
phase transition occurs (the condensation of the virtual molecular type
excitations with zero momentum and spin). The absorption typical for supercon
ducting metallic films below the metal plasma frequency appears. A
compressibility of the system ( energy capacity ) increases at this point.
Energy gap (1 - 1.4) eV at 200 GPa corresponds to a number of pairs about
0.2 number of atoms.
  Of course, all the problems that arise in discussion concerning metallic
xenon properties cannot be solved without direct experiments on magnetism.
>From the other side these experiments would be interesting taking into
account the existence of some virtual structures inside dielectric
condensates of IG atoms. May be they are simulate to some materials and
interactions discussed earlier in connection with the HTSC problem and
sometimes mentioned in search of adequate description of the contemporary
HTSC.

\newpage
\centerline{\bf\large References}
\begin{enumerate}
\item Mendelsson K.  The Quest for Absolute Zero.  World
University Library,  Weidenfeld  and  Nicolson . (1968).
\item Herzfeld K.F.  Phys Rev. {\bf29}, 701 (1927).
\item Wigner E., Hungtinton H.B. J. Chem. Phys.
{\bf3}, 764 (1935).
\item Goettel K.A., Eggert J.H., Silvera I.F., Moss W.C.
Phys. Rev. Lett. {\bf62}, 665 (1989).
\item Reichlin R.R., Brister K.E., McMahan A.K., Ross M..,
Martin S., Vohra Y.K., Ruof A.L.
Phys. Rev. Lett. {\bf62}, 669 (1989).
\item Ginsberg D.M., Tinkham M. Phys.Rev.
{\bf118}, 990 (1960).
\item Grant F. Rev. Mod. Phys. {\bf31}, 646 (1959).
\item Bogomolov V.N., Kudinov E.K. Firsov Yu.A.
Sov. Phys.-Sol. State. {\bf9}, 2502 (1968).
\item Bogomolov V.N. Phys. Solid State. {\bf35},469 (1993) (a);
Phys. Rev. {\bf51}, 17040 (1995)(b); Tech. Phys.Lett.
{\bf21}, 928 (1995) (c).
\item Bogomolov V.N. Preprint 1734 RAS. A.F.Ioffe
Phys.-Techn. Inst. (1999).
\end{enumerate}


\begin{figure}[tbp]
\epsfxsize=16cm
\epsfbox{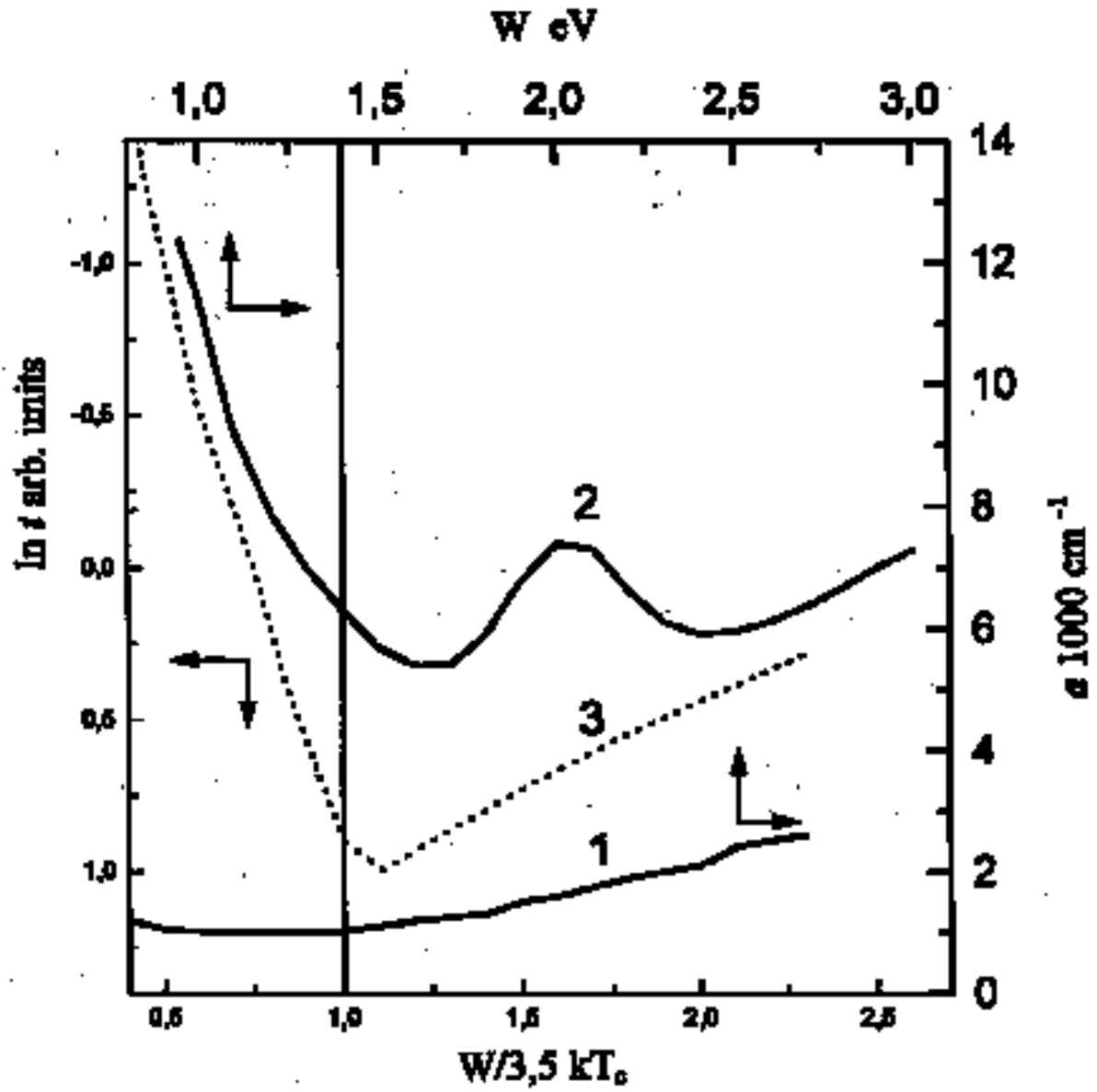}
\caption{The frequency dependence of the absorption coefficient
$a$ [ 4 ] (1)- at pressure 130 GPa ;  (2) - at pressure 200 GPa ;  (3) -
the calculated absorption coefficient  $\ln t$  of superconducting films
dependence on the relative photon energy $W/kT$  [ 6 ].}
\label{fig1}
\end{figure}

\begin{figure}[tbp]
\epsfysize=18cm
\epsfbox{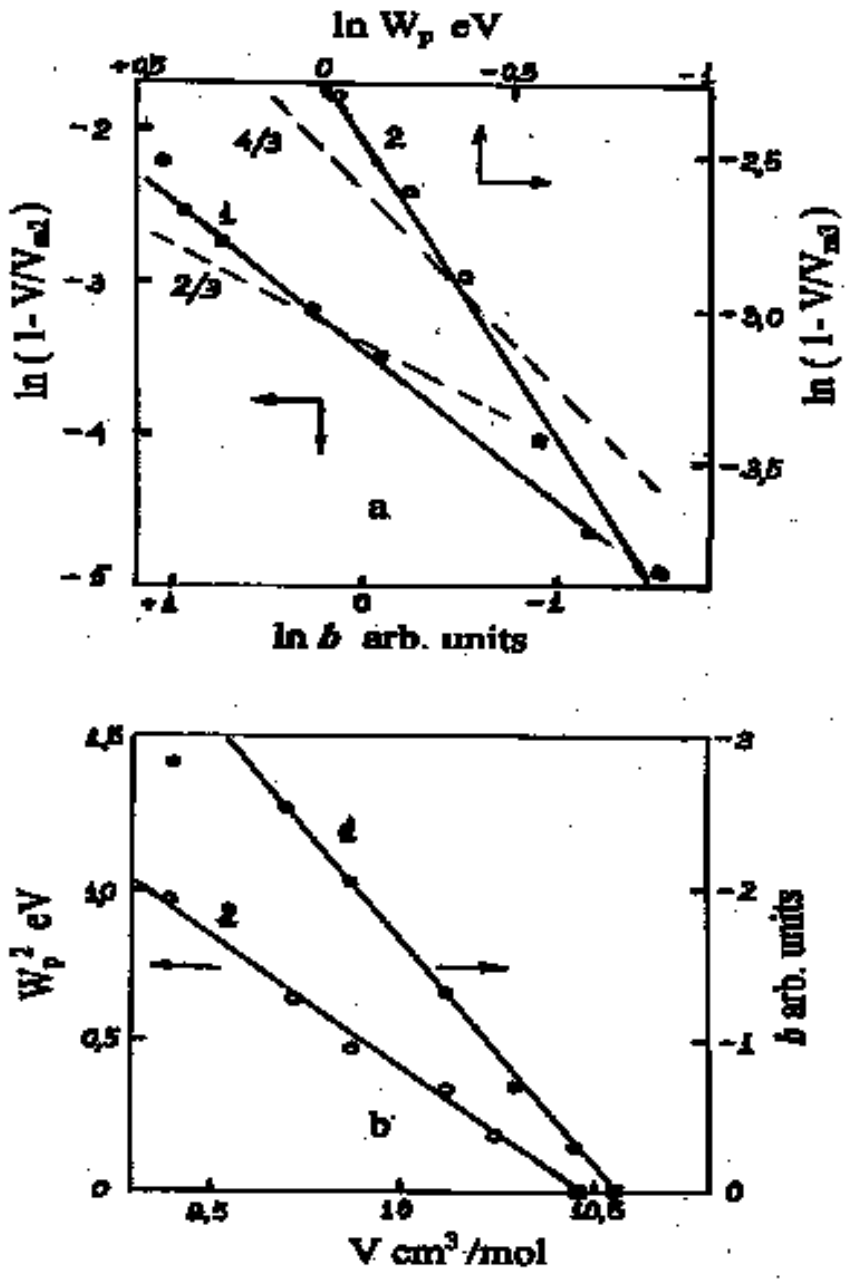}
\caption{a --- the logarithmic dependencies of  $W_p$ (eV) on $V/V_{m3}$
and of the absorption  b (arb. units) at 2 eV on
$V/V_{m2}$ [ 4 ]. The numbers at straight lines are powers in expressions
(1) and (2);  b --- the volume dependence of
$W_{p2}$ (eV ) and  $b$ (arb units). The numbers at straight lines are
powers in expressions (1) and (2).  Square points  are
$V_{m2} = 10.56$ cm$^3$ /mol and
$V_{m3} = 1046$ cm$^3$ /mol.}
\label{fig2}
\end{figure}

\begin{figure}[tbp]
\epsfxsize=16cm
\epsfbox{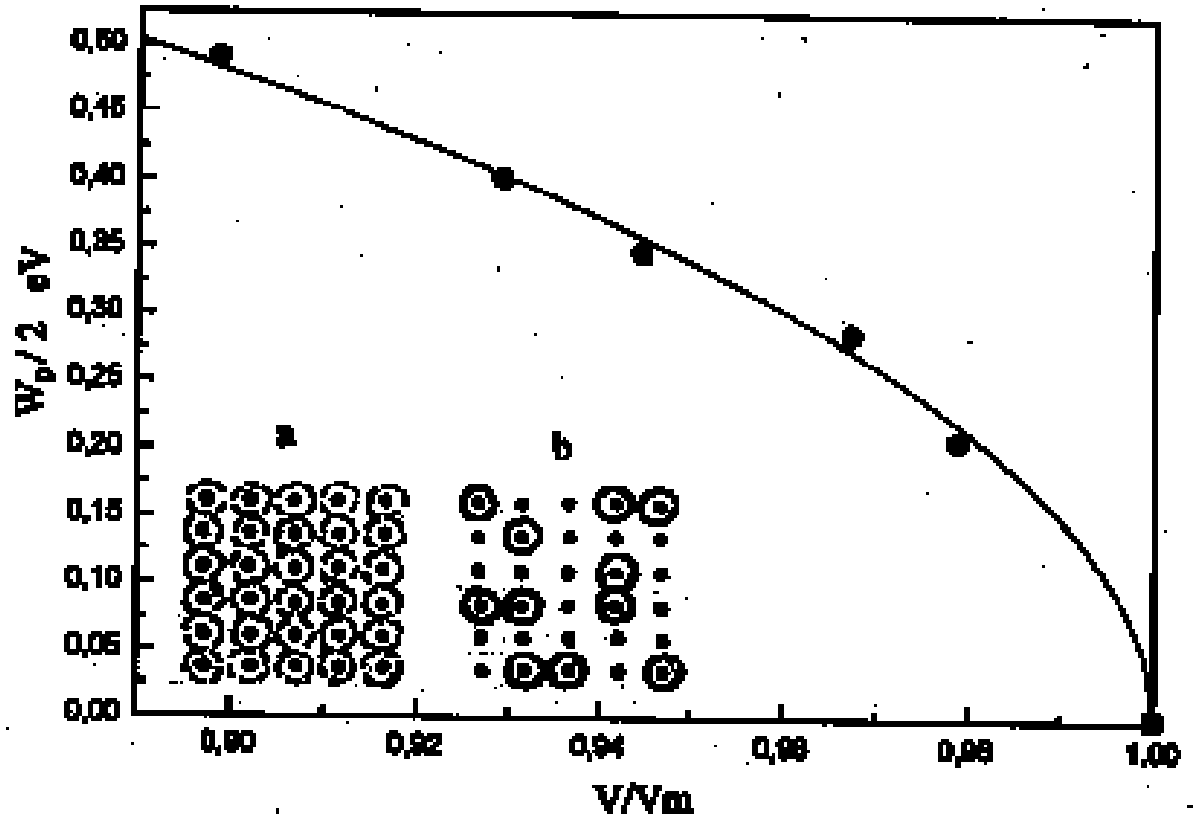}
\caption{The volume dependence of  $W_p$ (eV) at $A=0.48$ eV. The points
are the experimental data at 200 GPa [ 4 ].
Insertions - a --- the average electron density $eX$ distribution at
excited state orbitals in a lattice relaxation time scale;  b --- The
instantaneous ( momentary ) distribution of electrons at $X$ lattice sites
of condensate in the electron relaxation time scale.}
\label{fig3}
\end{figure}

\begin{figure}[tbp]
\epsfxsize=16cm
\epsfbox{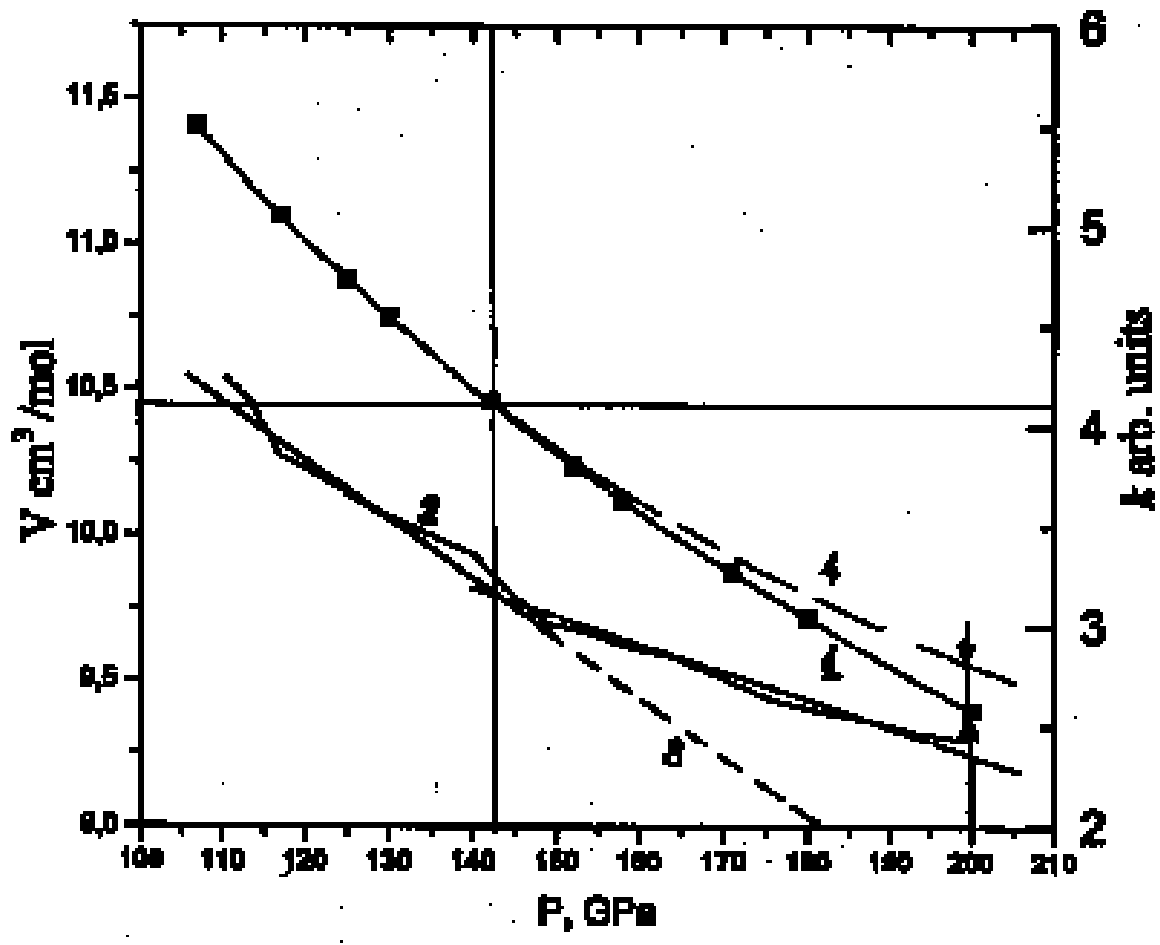}
\caption{The pressure dependence of the volume $V$ for xenon.
1 --- the points are the experimental data [ 4 ];
2 --- the compressibility  $k$. The straight lines are approximation;
3 --- the compressibility of the dielectric phase extrapolated into the
metallic phase region;
4 --- the $P-V$ dependence at the compressibility corresponding to the
dielectric phase. Arrows show the hypothetic transition into the metallic
phase from the curve 4 to the curve 1  at 200 GPa.}
\label{fig4}
\end{figure}

\begin{figure}[tbp]
\epsfxsize=16cm
\epsfbox{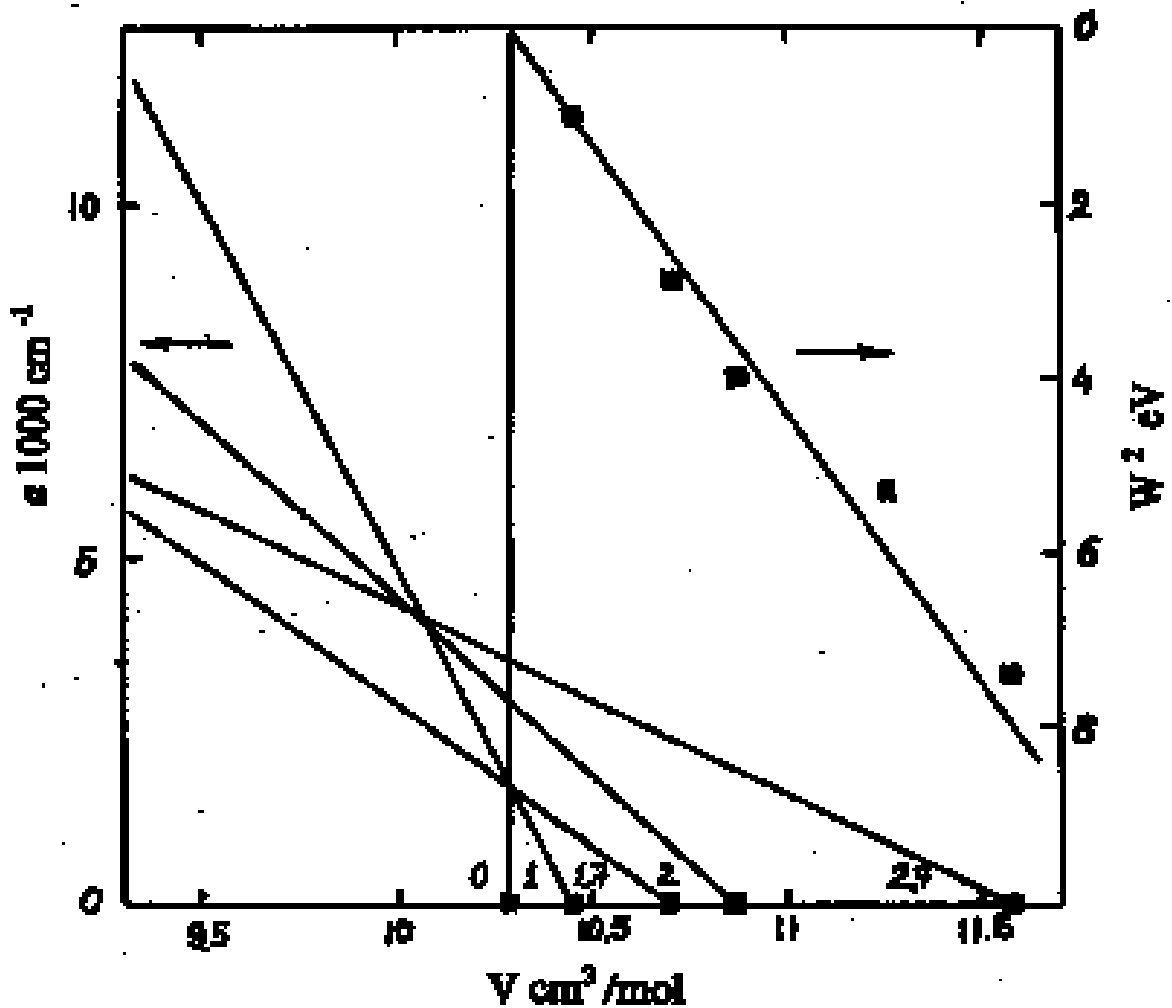}
\caption{The dependence of the light absorption coefficient on the Xe volume
[ 4] at the frequencies 1 , 1.7 , 2 , 2.7 eV ( the numbers at the straight
lines ).
a --- the dependence of the critical volumes $V_m$ (the points) on the
squared light frequency.}
\label{fig5}
\end{figure}

\end{document}